\documentstyle[10pt,a4]{article}
\begin{document}

ELECTRONIC  STRUCTURES OF QUANTUM DOTS AND

THE ULTIMATE RESOLUTION OF INTEGERS

\hspace{1.0in}

\hspace{1.0in}C.G. Bao, Y.Z. He, and G.M. Huang

Department of Physics, Zhongshan University, Guangzhou, 510275, China

\hspace{1.0in}

ABSTRACT: The orbital angular momentum L as an integer can be ultimately
factorized as a product of prime numbers. We show here a close relation
between the resolution of L and the classification of quantum states of an
N-electron 2-dimensional system. In this scheme, the states are in essence
classified into different types according to the $m(k)$-accessibility,
namely the ability to get access to symmetric geometric configurations. The $%
m(k)$-accessibility is an universal concept underlying all kinds of
2-dimensional systems with a center. Numerical calculations have been
performed to reveal the electronic structures of the states of different
types. This paper supports the Laughlin wave function and the composite
fermion model from the aspect of symmetry.

\hspace{1.0in}


{\bf Introduction, the resolution of integers and the accessibility of the
 configurations with a m-fold axis}

The resolution of integers is a basic and important theorem in the primary
theory of number. Each integer I can be ultimately factorized as a product
of prime numbers as I= $2^{n_2}3^{n_3}5^{n_5}7^{n_7}\cdot \cdot \cdot $.
Thus, like the elementary particles in physics, the prime numbers serve as
elementary elements of integers. Let the set of the idempotent indexes be
denoted as $\{n_i\}$ . Evidently, the character of an integer is determined
by this set. It is possible that the speciality of the set $\{n_i\}$ of an
integer would affect the role of the integer in nature. In quantum
mechanics, integers play a very important role. A number of physical
quantities are integers (if specific units are used), e.g., the number of
particles in a system, the orbital angular momenta L and their components,
the spin S, the charge Z , etc.. Therefore, there might be a connection
between the theory of number and quantum mechanics. However, such a
connection is not clear until now. In particular, how the idempotent series $%
\{n_i\}$ play their role directly in physical world is not clear. For three
dimensional systems there are magic integers ( e.g., the numbers 2, 8, 20,
28, 50, 82, $\cdot \cdot \cdot $ for the shell structures of nuclei; the
numbers 2,10, 18, 36, 54, $\cdot \cdot \cdot $ for the periodic table of
atoms, the numbers 13, 19, 25, 55, 71 $\cdot \cdot \cdot $ for the abundance
of atomic clusters, etc.). It is not clear whether these magic integers have
something special in their $\{n_i\}$ . Nonetheless, for two-dimensional
quantum dots, we show here a close relation between the set $\{n_i\}$ of L
as an integer and the electronic structures.

Let us consider a two-dimensional system of N electrons confined in a
quantum dot $^{1-3}$. N may be large or small, but is finite. It is assumed
that the potential of confinement is isotropic so that the orbital angular
momentum L together with the total spin S are conserved in an eigenstate $%
\Psi _{LS}$ . It is well known that the spin-states are the basis-states of
the two-row representation $\stackrel{\sim }{\lambda }$ of the permutation
group, $\stackrel{\sim }{\lambda }$ =[$\frac N2$+S,$\frac N2-$S] . Let the $%
i $-th spin-state of the $\stackrel{\sim }{\lambda }$ representation be
denoted as $\chi _i^{\stackrel{\sim }{\lambda }}$ , $i$ = 1 to $d$ ( the
dimension of $\stackrel{\sim }{\lambda }$ ). $\Psi _{LS}$ can be expanded as

$\Psi _{LS}=\sum_iF_i^\lambda \chi _i^{\stackrel{\sim }{\lambda }}$%
\hspace{1.0in}(1)

where $F_i^\lambda $ is a function of spatial coordinates and is a
basis-state of the $\lambda $ representation, the conjugate representation
of $\stackrel{\sim }{\lambda }$. . In (1) the antisymmetrization is assured.

On the other hand, when the number of electrons N or N-1 can be factorized
as a product of integers N=N$_A$N$_B$ or N$-$1=N$_C$N$_D$, the electrons may
surround the center of confinement and form a geometric configuration with a
m-fold axis, m=N$_A$, N$_B$, N$_C$, or N$_D$ . Such a configuration is
called a $m(k)$ configuration, where k=N/m or (N-1)/m is the number of
homocentric circles, each contains m electrons (some circles might have the
same radius). When k=(N-1)/m, the $m(k)$ configuration would have an
electron at the center. Some of the $m(k)$ configurations are in the domain
of low total potential energy, these $m(k)$ are important to the electronic
structures as we shall see. Since a rotation of a $m(k)$ about the center by 
$\frac{2\pi }m$ is equivalent to $k$ cyclic permutations of particles, we
have$^4$ at a $m(k)$

$e^{i2\pi L/m}F_i^\lambda (12\cdot \cdot \cdot \cdot )=F_i^\lambda (23\cdot
\cdot \cdot \cdot )=\sum_jG_{ji}^\lambda (p_c)F_j^\lambda (12\cdot \cdot
\cdot \cdot )$\hspace{1.0in}(2)

where $G_{ji}^\lambda (p_c)$ is the matrix element of the $\lambda $%
-representation associated with the $k$ cyclic permutations $p_c$ . There
are totally $d$ such equations, they form a set of homogeneous linear
equations. From the set we define a determinant

$D(L,\lambda ,m)=|G_{ji}^\lambda (p_c)-\delta _{ij}e^{i2\pi L/m}|$%
\hspace{1.0in}(3)

Evidently, if $D(L,\lambda ,m)$ is nonzero, the set of linear equations will
have only zero solutions, and thereby the $F_i^\lambda $ must all be zero at
the $m(k)$. In this case, an inherent nodal surface is imposed by symmetry
and the $m(k)$ is therefore inaccessible$^5$. If a wave function is
distributed in a domain containing an inaccessible configuration, the
inherent nodal surface would cause an excited oscillation resulting in a
great increase in energy. Hence, for low-lying states, the wave function
would be far away from the inaccessible configuration. Anyway, whether $%
D(L,\lambda ,m)$ is nonzero or zero would affect strongly the electronic
structure of the state.

Since $D(L,\lambda ,m)$ depends on L and S, evidently the electronic
structures depend strongly on L and S. The calculation of $D(L,\lambda ,m)$
is not difficult if N is small. However, a general discussion is not easy
due to the complexity in the general representation of {\it S}$_N$ group.
Nonetheless, for polarized systems, the discussion becomes much simpler as
follows.

Let the group of states having the same $L$ be called a $L-series$. For
polarized systems we have S=N/2 and $\lambda $ is totally-antisymmetric. In
this case the discriminant reads

$D(L,\lambda ,m)=(-1)^{(m-1)k}-e^{i2\pi L/m}=0$\hspace{1.0in}(4)

If a couple of $m$ and $k$ fulfil (4), then the corresponding $m(k)$ is
accessible to the $L-series$. Evidently, all the $m(k)$ are accessible to
the L=0 series except the case of m even and k odd. Furthermore, all the $%
m(k)$ are inaccessible to the L=1 series except the case of m=2 and k odd.
When $L\geq 2$, from the resolution of integers, we have

$L=2^{n_2}3^{n_3}5^{n_5}7^{n_7}(11)^{n_{11}}\cdot \cdot \cdot \cdot $%
\hspace{1.0in}(5.1)

$m=2^{m_2}3^{m_3}5^{m_5}7^{m_7}(11)^{m_{11}}\cdot \cdot \cdot \cdot $%
\hspace{1.0in}(5.2)

Inserting (5) into (4), we have

$\exp [i\pi (2^{n_2+1-m_2}3^{n_3-m_3}5^{n_5-m_5}\cdot \cdot \cdot
)]=(-1)^{(m-1)k}$\hspace{1.0in}(6)

From (6) and by using a little primary knowledge of the theory of number we
arrive at the following rules:

RULE 1, If $m$ is odd, or if $m$ and $k$ are both even, then the $m(k)$ is
accessible to the $L-series$ with $n_i\geq m_i$ (here $i=2,3,5,7,\cdot \cdot
\cdot $) .

RULE 2, If $m$ is even and $k$ is odd, then the $m(k)$ is accessible to the $%
L-series$ with $n_2=m_2-1$ and $n_i\geq m_i$ .

RULE 3, Let $mk=m^{\prime }k^{\prime }=I$ . If the integers $m$ and $%
m^{\prime }$ do not have a common factor, and if both the $m(k)$ and $%
m^{\prime }(k^{\prime })$ are accessible to a $L-series$ , then the
product-configuration $mm^{\prime }(\frac I{mm^{\prime }})$ is also
accessible to the $L-series$ .

RULE 4, Let $mk=m^{\prime }k^{\prime }=I$ . If the configuration $mm^{\prime
}(\frac I{mm^{\prime }})$ is accessible to a $L-series$, then both the $m(k)$
and $m^{\prime }(k^{\prime })$ are accessible to the $L-series$ .
Alternatively, if $m(k)$ (or $m^{\prime }(k^{\prime })$ ) is inaccessible to
a $L-series$ , the $mm^{\prime }(\frac I{mm^{\prime }})$ is also
inaccessible to the $L-series$ .

RULE 5, If $_{}L=mk(mk\pm j_o)/2\geq 0$, where $j_o$ is an odd integer ,
then the $m(k)$ is accessible to the $L-series$.

For examples, (i) The 8(3) configuration of the N=24 or 25 system has $m_2=3$
and $m_i=0$ ($i\geq $3). According to RULE 2, this configuration is
accessible to the L=4$j_o$ series, here $j_o$ is an arbitrary positive odd
integer. (ii) According to RULE 1 the 3(6) is accessible to the L=3$j$
series, here $j$ is an arbitrary positive integer. According to RULE 2 the
2(9) is accessible to the L=$j_o$ series. Therefore both the 3(6) and 2(9)
are accessible to the L=3$j_o$ series. Since 3 and 2 do not have a common
factor, according to RULE 3, the 6(3) is also accessible to the L=3$j_o$
series. (iii) Since the 8(1)\ is accessible to the L=4$j_o$ series,
according to RULE 4 both the 4(2) and 2(4) are also accessible to the
series. Alternatively, since the 2(4)\ is inaccessible to the L=$j_o$
series, according to RULE 4 both the 4(2) and 8(1) are also inaccessible to
the series.

It is straight forward to know from the RULE\ 5 that all the $m(k)$ with $%
mk= $N are accessible to the $L=$N(N-1)/2 $\pm j$ N states, and all the $%
m(k) $ with $mk=$N-1 are accessible to the $L=$N(N-1)/2 $\pm $ $j$ (N-1)
states . Therefore, all the $m(k)$ disregarding $mk=$N or N-1 are accessible
to the $L=j_o$N(N-1)/2 states, i.e., these special states do not contain
inherent nodal surfaces at any $m(k)$, therefore they are specially stable.
When a magnetic field is applied, they are the strongest candidates of
ground states. Here the reciprocal of $j_o$ is associated with the filling
factor $\nu $ of the Hall effect$^{6,7}$, and the $L=j_o$N(N-1)/2 states are
associated with $\nu =1,\frac 13,\frac 15,\cdot \cdot \cdot $

It is recalled that the famous Laughlin wave function$^{8,9}$.

$\psi _\nu =[\Pi _{i<j}(z_i-z_j)^{j_o^{\prime }}]\exp (-\sum_jz_j^{*}z_j)$%
\hspace{1.0in}(7)

has $L=j_o^{\prime }$N(N-1)/2 , here $j_o^{\prime }$ is also an odd integer.
Thus, we have proved that this state does not contain inherent nodal
surfaces at any $m(k)$ configuration, therefore the wave function can be
smoothly distributed without nodes in the domain of low potential energy.
This might be a reason that they are close to the exact solutions.

\hspace{1.0in}

{\bf Classification of quantum states in a 9-electron dot}

We shall see that the accessibility of the $m(k)$-configurations affects the
electronic structures greatly. Let us investigate in detail a 9-electron
dot. Although such a system has already been more or less concerned in the
literatures$^{10-13}$, a precise calculation beyond the lowest Landau level
approximation and a detailed analysis of the wave functions have not yet
been performed.

In the view of geometry there are the 9(1), 3(3), 8(1), 4(2), and 2(4)
configurations. However, due to the RULE\ 1 and 2, only some of them are
accessible to a specific $L-series$ . Furthermore, the 9(1) can be neglected
due to having a much higher potential energy ( In 9(1) only nine bonds can
be optimized, while in 8(1) sixteen bonds can be. Incidentally, the 9(1)
would become more important in a 10-electron system, in that case eighteen
bonds can be optimized). Thus the important $m(k)$ are the other four, they
lie in the domain of lower potential energy. Based on their accessibility,
the $L-series$ can be classified into eight types as shown in TABLE 1. Due
to the RULE\ 1 and 2, the scheme depends straight on $\{n_i\}$ . For an
example, the $L-series$ having $n_2=2$ and $n_3\geq 1$ are both 8(1)- and
3(3)-accessible. According to the RULE 4, the 4(2), and 2(4) are also
accessible to this series. Thus they are inherently nodeless in all the
important $m(k)$, and are grouped to type 1. Consequently, they are superior
in stability and therefore particularly important. They have $L=j_o$%
N(N-1)/6, thus the above mentioned Laughlin states ($L=j_o^{\prime }$%
N(N-1)/2) are members of this type. The other members of this type are also
candidates of the ground state and are associated with the filling factor $%
\nu =3/j_o$ .

For another example, the type 4 is 3(3)-accessible but 2(4)-, 4(2)- and
8(1)-inaccessible. Incidentally, the $n_i$ with $i\geq 5$ are irrelevant to
the classification of the 9-electron system and therefore can be arbitrary.

TABLE 1, Classification of states of a polarized 9-electron dot according to
the $\{n_i\}$ of L. The $m(k)$ configurations accessible to a specific type
are listed.

\begin{tabular}{|ccc|}
\hline
\multicolumn{1}{|c|}{Type} & \multicolumn{1}{c|}{accessible $m(k)$} & $%
\{n_i\}$ \\ \hline
\multicolumn{1}{|c|}{1} & \multicolumn{1}{c|}{3(3),8(1)} & $n_2=2,n_3\geq 1$
\\ \hline
\multicolumn{1}{|c|}{2} & \multicolumn{1}{c|}{3(3),4(2)} & $n_2\geq
3,n_3\geq 1$ \\ \hline
\multicolumn{1}{|c|}{3} & \multicolumn{1}{c|}{3(3),2(4)} & $n_2=1,n_3\geq 1$
\\ \hline
\multicolumn{1}{|c|}{4} & \multicolumn{1}{c|}{3(3)} & $n_2=0,n_3\geq 1$ \\ 
\hline
\multicolumn{1}{|c|}{5} & \multicolumn{1}{c|}{8(1)} & $n_2=2,n_3=0$ \\ \hline
\multicolumn{1}{|c|}{6} & \multicolumn{1}{c|}{4(2)} & $n_2\geq 3,n_3=0$ \\ 
\hline
\multicolumn{1}{|c|}{7} & \multicolumn{1}{c|}{2(4)} & $n_2=1,n_3=0$ \\ \hline
\multicolumn{1}{|c|}{8} & \multicolumn{1}{c|}{} & $n_2=0,n_3=0$ \\ \hline
\end{tabular}

\hspace{1.0in}

The classification according to $\{n_i\}$ is in essence a classification
according to the accessibility of the $m(k)$ , or in other words according
to the inherent nodal surfaces. Thus the classification is model-independent
and based simply on the fundamental principle of symmetry. To show the
reasonableness of the classification, numerical results are given in the
follows.

The Hamiltonian reads

$H=\sum_i[\frac{p_i^2}{2m*}+\frac 12m^{*}\omega _o^2r_i^2]+\frac{e^2}{4\pi
\varepsilon }\sum_{i>j}\frac 1{r_{ij}}$\hspace{1.0in}(8)

where $m^{*}$ is the effective mass, $\varepsilon $ the dielectric constant.
It is assumed that $m^{*}=0.067m_e$ , $\varepsilon =12.4$ (for GaAs dots),
and $\hbar \omega _o$ = 3 meV. In what follows meV and $\sqrt{\hbar
/(m^{*}\omega _o)}$ will be used as units for energy and length,
respectively.

Let the single-electron harmonic oscillator states be denoted as $%
|ll^{\prime },\omega \rangle ,$ they have energies $(l+l^{\prime }+1)\hbar
\omega $ and angular momenta $l-l^{\prime }$ . With them antisymmetrized
harmonic oscillator product states $\Phi _J=$ $|l_1l_1^{\prime
},l_2l_2^{\prime },\cdot \cdot \cdot l_Nl_N^{\prime },\omega \rangle $ with $%
\sum_i(l_i-l_i^{\prime })=L$ are constituted and are used as basis functions
of eigenstates. Here $\omega $ is in general not equal to $\omega _o$ , but
is adjustable to minimize the eigenenergies. For all the following
calculations we have $l_i$ $\leq $ $25$ , $\sum_il_i^{\prime }\leq $ $3$ ,
i.e., higher Landau levels are included. In order to depress the number of
basis functions, $\Phi _J$ are arranged in such a way so that $<\Phi
_J|H|\Phi _J>\leq <\Phi _{J+1}|H|\Phi _{J+1}>$ . In such a sequence the one
with a very large index $J$ is not important to the low-lying states. Then, $%
H$ is diagonalized first in a space with $J$ starting from 1 to a given
smaller number, and again to a larger number, and repeatedly, until a
satisfied convergency is achieved, i.e., the eigenenergies have at least
four effective figures and the correlated densities extracted from the
related wave functions are nearly unchanged. It was found that, even in the
case of N=19, $J\leq 8000$ is enough for our purpose if the variational
parameter $\omega $ has been properly adopted and if L is not much larger
than N(N-1)/2 (e.g., L $\leq$ 100 if N=9).

Once an eigenstate $\Psi _{LS}$ is obtained, the associated 1-body, 2-body,
and 3-body density functions $\rho _1$, $\rho _2$ and $\rho _3$ would be
extracted (the results of $\rho _1$ will not be given here)$.$ For example,
the 3-body density function is defined as

$\rho _3(\stackrel{\rightarrow }{r}_1,\stackrel{\rightarrow }{r}_2,\stackrel{%
\rightarrow }{r}_3)=\int d\stackrel{\rightarrow }{r}_4\cdot \cdot \cdot d%
\stackrel{\rightarrow }{r}_9|\Psi _{LS}|^2$\hspace{1.0in}(9)

It has been previously suggested$^{10-12}$ that some of the electrons ( N$%
_{out}$ ) might be located in a ring outside, and form a N$_{out}$%
-ring-structure. For a 9-electron dot, it turns out that the total potential
energy of a ring-structure with N$_{out}\leq 4$ or N$_{out}$=9 is much
higher. The domain in coordinate space containing the 5- to
8-ring-structures are broad, where the total potential energy is low and
flat . If symmetry is not taken into account, these ring-structures might be
equally preferred by low-lying states. However, if the domain of a
ring-structure contains an inaccessible $m(k)$ (e.g., the domain of a 6-ring
structure contains the 3(3) which is inaccessible to the type 5 to 8), the
ring-structure would be unfavorable because the existing inherent nodal
surface would cause a great increase in energy. Thus, which ring-structures
would be the better choice depends on the type of states.

Let the $n$-th state and its energy of a $L-series$ be denoted as (L)$_n$
and E$_n$(L) . The $n$=1 state, namely the lowest of the series, is called
the first-state. Let us define

E$_{cusp}(L)=$E$_1$(L)-(E$_1$(L-1)+E$_1$(L+1))/2\hspace{1.0in}(10)

Evidently, if E$_{cusp}(L)$ is negative, the $L-series$ would have a
downward cusp.

Now, let us first inspect the L=60 series with $\nu $=N(N-1)/2L=3/5 as an
example of type 1. We have E$_1$(60)=272.86 and E$_{cusp}(60)=$ -0.43. Thus,
the L=60 series has a downward cusp, a common feature of the type 1. The
2-body densities $\rho _2$ of the (60)$_1$ and (60)$_2$ are plotted in
Fig.1a and 1b, where the 8-ring and 6-ring structures originate from the
8(1)- and 3(3)- accessibility, respectively. Since the 8-ring has more
electrons in the ring, its moment of inertia is larger resulting in having a
smaller rotation energy. Therefore it is lower than the 6-ring in the case
of L=60.

The L=36 series belongs to the type 1 with $\nu $=1. The $\nu $=1 states are
special because they have only one basis function in the lowest Landau
level, namely the Laughlin wave function (eq.(7)) with $j_o^{\prime }=1$ .
Therefore the (36)$_1$ will be dominated by this function. In fact, in our
calculation, this state has the weights of the lowest to the fourth lowest
Landau levels to be 80.0\%, 16.8\%, 2.5\%, and 0.7\%, respectively. It is
noted that the clear geometric features shown in Fig.1a and 1b arise from a
coherent mixing of the basis functions. Although the (36)$_1$ is allowed by
symmetry to get access to symmetric geometric configurations, this state is
not able to possess a clear geometric feature as shown in Fig.1c due to the
lack of coherent mixing. In fact, the feature of Fig.1c arises simply from
the Laughlin wave function. Incidentally, the (36)$_1$ has a rather low
energy E$_1$(36)=209.18 and a very large gap 4.98 lying between E$_1$(36)
and E$_2$(36), thus this state is superior in stability.

Fig.1d to 1f are examples of type 4. They do not have the 8-ring structure
because this type is 8(1)-inaccessible. The (63)$_1$ and (81)$_1$ have a
clear 6-ring originating from the 3(3) accessibility. To see clearer the
structure of the core, the $\rho _3$ of the (81)$_1$ is plotted in Fig.2a,
where a clear regular triangle is inside. However, instead of having a
6-ring, the (99)$_1$ has a 7-ring structure. It is noted that a 9-particle
system does not contain the 7(k)-configuration. Hence, the 7-ring is not
constrained by symmetry. Thus, it is not surprising that both the 6- and
7-ring emerge in the type 4. The 6-ring would be better than the 7-ring if L
is smaller (e.g., the (63)$_2$ and (81)$_2$ are found to have a 7-ring ).
However, the 7-ring would be better if L is larger due to having a larger
moment of inertia.

It is clear that , although the inherent nodal surfaces have imposed serious
constraints on wave functions, the electronic structures are not uniquely
determined by them. In addition to the pairwise interaction, the centrifugal
barrier and the parabolic potential also play their role. The barrier leads
to the preference for the configurations with a larger moment of inertia.
Therefore, a critical value(s) of L denoted as L$_{crit}$ might exist for
each type so that the first-states with L smaller than L$_{crit}$ and those
with L$\geq $L$_{crit}$ are distinct in structure (e.g., the (81)$_1$ and
(99)$_1$). The parabolic potential confines the number of effective basis
functions taking part in coherent mixing. Thus the lower states with $\nu $
equal or close to 1 are insufficient in coherent mixing and therefore
ambiguous in geometric feature (e.g., the (36)$_1$).

In addition to the case of $\nu $ =1, an example of $\nu $ =18/19 is shown
in Fig.1g to 1i belonging to type 7. The L=38 states have only two basis
functions in the lowest Landau level. Consequently, both the (38)$_1$ and
(38)$_2$ are ambiguous in geometric feature. However, the (38)$_3$ dominated
by the second lowest levels has a clear 7-ring due to having a sufficient
coherent mixing. There is a very large gap 3.15 lying between the E$_2$(38)\
and E$_3$(38). The two lower states have E$_1$(38)=220.61 and E$_2$%
(38)=221.25, much higher than the E$_1$(36). Noting that the type 7 is both
8(1)- and 3(3)-inaccessible. Therefore , the 7-ring is preferred.. Another
example of type 7 is given in Fig. 1j. Since a number of the first Landau
levels are contained in the L=50 series, instead of the third-state, the
7-ring appear first in the first-state.

Fig.1k and 1$l$ show the similarity of the two first-states of type 2, both
have a 6-ring. Fig.1m and 1n show the similarity of the two second-states,
both have a square-structure. Evidently, these structures originates from
the 3(3)- and 4(2)-accessibility. To see clearer the square, the $\rho _3$
of the (48)$_2$ is plotted in Fig.2b. Since these states are
8(1)-inaccessible, the inherent nodal surfaces at the 8(1) would spoil the
stability of the square. Thus the square-structure is higher.

Fig.1o and 1p are examples of type 5. Since this type is 8(1)-accessible but
3(3)-inaccessible, it is easy to understand why the octagon shape emerges.

\hspace{1.0in}

{\bf Classification of fermion states of N}$\neq ${\bf 9 dots}

For the classification of states of a general N-electron dot, we have first
to figure out which $m(k)$ configurations with $mk$=N or N-1 will be
contained in the domain of lower potential energy. Secondly, we have to make
sure their accessibility to the $L-series$ .

If N=6, the 5(1), 3(2), 2(3) configurations should be considered (the 6(1)
is automatically taken into account due to the RULE 3). Then the type 1 has
L=$j_o$15, which is the intersection of \{L$\equiv $0 mod 5\}and \{L$\equiv $%
3 mod 6\}. This is a well known result$^{14}$.

When N is larger, the effect of the $m(k)-$accessibility might reduce,
because in the coordinate space the domain of low energy is so broad that
the wave function is easy to avoid the inaccessible configuration. However,
even if N\ is as large as 19, the effect of the $m(k)$-accessibility is
still explicit. When N=19, the 9(2), 3(6) and 2(9) have to be considered
(the 6(3) is automatically taken into account due to the RULE 3). Then the
classification is shown in TABLE 2.

TABLE 2, Classification of states of a polarized 19-electron dot.

\begin{tabular}{|ccc|}
Type & accessible $m(k)$ & $\{n_i\}$ \\ 
1 & 9(2),3(6),2(9) & $n_2=0,n_3\geq 2$ \\ 
2 & 3(6),2(9) & $n_2=0,n_3=1$ \\ 
3 & 9(2) & $n_2\geq 1,n_3\geq 2$ \\ 
4 & 3(6) & $n_2\geq 1,n_3=1$ \\ 
5 & 2(9) & $n_2=0,n_3=0$ \\ 
6 &  & $n_2\geq 1,n_3=0$ \\ 
&  & 
\end{tabular}

It was found that both the type 1 and type 2 are better in stability, they
have downward cusps. For examples, we have E$_{cusp}$(177)=$-0.032$ and E$%
_{cusp}$(183)=$-0.014$, both belong to type 2. Fig.2c show a 12-ring
structure originating from the 6(3)-accessibility (similar to the fact that
the 6-ring structure originates from the 3(3)-accessibility, cf. Fig.2a ).
this 12-ring structure is common to the type 1 and 2. On the other hand, the
geometric feature of the (184)$_1$ of type 6 is not very clear, but the
ring-structure is explicit ( 13 electrons are found in the ring ). These
figures support the ring structure proposed by other authors$^{10-12}$.

In general, not matter how large N is, only the states with a superior
stability are interesting. The stability is quite often associated with the
geometric symmetry. Once the geometric symmetry is concerned, the effect of
the $m(k)$-accessibility should be considered.

\hspace{1.0in}

{\bf Bosonic systems}

Since the above discussion is model-independent and is simply based on
symmetry consideration, it can be generalized to bosonic systems as well. In
this case the wave functions should be completely symmetric with respect to
particle permutation. Thus, instead of eq.(4), we have the criterion

$1-e^{i2\pi L/m}=0$\hspace{1.0in}(10)

and accordingly the $m(k)$ configuration is accessible to the $L-series$
with $n_i\geq m_i$ disregarding $m$ and $k$ are even or odd. In particular,
all the $m\left( k\right) $ are accessible to the L=0 states.

In recent years the Bose-Einstein condensation has been extensively studied,
and the trapped atoms gases have been shown to Bose condense$^{15,16}$.
Cooper and Wilkin have studied the properties of rotating Bose-Einstein
condensates in parabolic traps. When the rotation frequency is larger, they
found the ground state angular momentum L for N=3 to 10- boson systems are
6, 12, 20, 30, 42, 56, 72, and 90, respectively$^{17}$. This can be
explained based on the composite fermion model$^{17,18}.$ Alternatively, we
now provide a model-independent explanation simply based on the $m(k)$%
-accessibility. For an example, the important $m(k)$ for the N=8 system is
7(1) and 4(2) (here the 8(1) is much less important than the 7(1), the
former has only eight bonds to be optimized while the latter has fourteen )
, thus the type 1 has $n_2\geq 2$ and $n_7\geq 1$. Therefore the L=56 state
would appear as a ground state when the rotation frequency lies in a
specific region. The important $m(k)$ for the N=9 system is 8(1) and 3(3),
thus the type 1 has $n_2\geq 3$ and $n_3\geq 1$, therefore the L=72 state
would appear as a ground state. The important $m(k)$ for the N=10 system is
9(1), 5(2), 3(3), and 2(5), thus the type 1 has $n_2\geq 1$ , $n_3\geq 2,$
and $n_5\geq 1$, therefore the L=90 state would appear as a ground state.

In fact, the downward cusps found in the ref.17 are closely related to the $%
m(k)$-accessibility. For examples, from the TABLE\ I of the ref.17 , we know
that the N=6 system has a cusp at L=6 and 12 which are associated with the
3(2) and 2(3) accessibility, a cusp at L=10 which is associated with the
5(1) and 2(3) accessibility, a cusp at L=15 which is associated with the
5(1) and 3(2) accessibility, etc.

\hspace{1.0in}

{\bf Concluding remarks}

An exact diagonalization of the Hamiltonian has been performed. Since higher
Landau levels have been considered, the numerical results (even in the case
of N=19) are very accurate in the qualitative sense. Since the $\omega $ of
the basis functions is a variational parameter, the convergency is thereby
greatly improved. Furthermore, in addition to the usually given $\rho _1$
and $\rho _2$ , $\rho _3$ have also been calculated to help the analysis.

A scheme of classification according to the idempotent series $\{n_i\}$ of L
has been proposed. In this scheme each type has its own $m(k)$%
-accessibility, or its own inherent nodal surfaces. The classification is
objective and model-independent. Although the electronic structures are not
uniquely determined by the $m(k)$-accessibility, its great effect has been
confirmed by the numerical results. Since the type 1 is inherently nodeless
in the domain of low potential energy, the first-states of this type are
superior in stability and are the strongest candidates of the ground states.
These noticeable states can be easily identified in our scheme.

The analysis of this paper supports the Laughlin wave function and the
composite fermion model from the aspect of symmetry.

The $m(k)$-accessibility is an universal concept for all kinds of
2-dimensional systems with a center. The introduction of this concept would
lead to a better understanding of these systems.

The theory of number and quantum mechanics are two previously unrelated
areas of science. Here we show a direct relation between the ultimate
resolution of L\ as an integer and the accessibility of the $m(k)$
configuration. This finding might lead to a closer relation between these
two important areas of science.

\hspace{1.0in}

Acknowledgment: This work is supported by the National Natural Science
Foundation of China.

\hspace{1.0in}

1, L. Jacak, P. Hawrylak, A. W\'ojs, {\it Quantum Dots }(Springer, Berlin,
1998)

2, T. Chakraborty, {\it \ Quantum Dots} (Elsevier, Amsterdam, 1999){\it \ }

3, G.W. Bryant, {\it \ Phys. Rev. Lett}. {\bf 59}, 1140 (1987)

4, C.G. Bao, {\it \ Phys. Rev. Lett}. {\bf 79}, 3475 (1997)

5, C.G. Bao, {\it Few-Body Systems}, {\bf 13}, 41 (1992)

6, K.V. Klitzing, G. Dorda, M. Pepper, {\it Phys. Rev. Lett.},{\bf \ 45},
494 (1980)

7, D.C. Tsui , H.L. St\"ormer, A.C. Gossard, {\it Phys. Reev. Lett.}, {\bf 48%
}, 1559(1982)

8, R.B. Laughlin, {\it \ Phys. Rev}. {\bf B23}, 5632 (1981)

9, R.B. Laughlin, {\it Phys. Rev. Lett.} {\bf 50}, 1395 (1983)

10, C. de C. Chamon, X.G. Wen, {\it Phys. Rev}. {\bf B49}, 8227 (1994)

11, H.M. Muller, S.E. Koonin, {\it Phys. Rev}. {\bf \ B54}, 14532 (1996)

12, E. Goldmann, S.R. Renn, {\it \ Phys. Rev.} {\bf B60}, 16611 (1999)

13, W.Y. Ruan, K.S. Chan, H.P. Ho, E.Y. B. Pun, {\it J. Phys.: Condens.
Matter} {\bf 12}, 3911 (2000)

14, P.A. Maksym, H. Imamura, G.P. Mallon, H. Aoki, {\it J. Phys.: Condens.
Matter}{\bf \ 12}, R299 (2000)

15, M.H. Anderson et al., {\it Science} {\bf 269}, 198 (1995)

16, F. Dalfovo, S. Giorgini, L.P. Pitaevskii, S. Stringari,{\it \ Rev. Mod
Phys}. {\bf 71}, 463 (1999)

17, N.R. Cooper, N.K.Wilkin, {\it Phys. Rev}. {\bf B60}. R16279 (1999)

18, J.K. Jain, T. Kawamura, {\it Europhys. Lett}. {\bf 29}, 321 (1995)

\hspace{1.0in}

\newpage\ 

FIGURE CAPTIONS

Fig.1, Contour plots of $\rho _2$ of a 9-electron system. The given electron
is marked by a black spot, its distance from the origin is given(in the unit 
$\sqrt{\hbar /(m^{*}\omega _o)}=194.7\AA $ ). This marked distance serves as
a scale for both the X and Y\ directions (slightly different scales have
been used for distinct states). The inmost contour (associated with the
highest peak) is marked by a double-line.

Fig.2, Contour plots of $\rho _3$ of 9- and 19-electron systems. Different
scales have been used for different states.

\end{document}